\input harvmac.tex
\input epsf
\input amssym.def
\input amssym

\def\capt#1{\narrower{
\baselineskip=14pt plus 1pt minus 1pt #1}}

\lref\Fa{ Faddeev, L.D. and  Takhtajan, L.A.: 
Hamiltonian Method in the
Theory of Solitons. New York: Springer 1987}

\lref\zama{Zamolodchikov, Al.B.: 
On the Thermodynamic Bethe Ansatz equation in sinh-Gordon
model. To appear}

\lref\tzev{Tsvelik, A.M.: Quantum Field Theory and
Condensed Matter Physics, Cambridge Univ. Press 1995 }

\lref\Leb{Lebedev, D. and Krachev, S.:
Integral representation for the eigenfunctions 
of quantum periodic Toda chain. Preprint ITEP-TH-54/99 
(hep-th/9911016) }

\lref\Gros{Gross, D. and Migdal, A.: Nonperturbative 
two-dimensional Quantum Gravity.
Phys. Rev. Lett. {\bf 64}, 127-130 (1990) }

\lref\Briz{Brezin, E. and Kazakov, V.: Exactly Solvable Field
Theories of Closed strings.
Phys. Lett. {\bf B236}, 144-150 (1990) }

\lref\Dug{Douglas, M. and Shenker S.: Strings 
in less than one-dimension. 
Nucl. Phys. {\bf B335}, 635-654 (1990) }

\lref\ziy{Seiberg, N.: Notes on Quantum Liouville Theory and 
Quantum Gravity. Prog. Theor. Phys. Supp. 
{\bf 102}, 319-349 (1990) }

\lref\cortt{ Curtright, T. and Thorn, C.: Conformally 
invariant quantization of the Liouville Theory. 
Phys. Rev. Lett. {\bf 48}, 
1309-1313 (1982)\semi
Braaten, E., Curtright, T. and Thorn, C.: 
An exact operator solution of the Quantum Liouville Field Theory.
Ann. Phys. {\bf 147}, 365-416 (1983)}

\lref\sklynn{Sklyanin, E.K.: Separation of variables. New Trends.
Preprint UTMS 95-9  (solv-int/9504001)}

\lref\dorn{Dorn, H. and Otto, H.-J.: 
On correlation functions for noncritical strings with
$c\leq 1,  D\geq 1$. Phys. Lett. {\bf B291}, 39-43 
(1992)\semi
Two and three point functions in Liouville Theory.
Nucl. Phys. {\bf B429}, 375-388 (1994)}

\lref\Zarn{Zamolodchikov, Al.B.: Mass scale
in the Sine-Gordon model and its
reductions. Int. J. Mod. Phys. {\bf A10}, 1125-1150  (1995) }

\lref\Al{Zamolodchikov, Al.B.: unpublished}

\lref\FLZZZ{Fateev, V., Lukyanov, S., Zamolodchikov, A. and
Zamolodchikov, Al.: Expectation values of local fields
in Bullough-Dodd model and integrable
perturbed conformal field theories.
Nucl. Phys. {\bf B516}, 652-674 (1998)}

\lref\Zar{Zamolodchikov, Al.B.: Thermodynamic Bethe ansatz in
relativistic models: Scaling 3-state Potts and Lee-Yang models.
Nucl. Phys. {\bf B342}, 695-720 (1990)}

\lref\Zarr{Zamolodchikov, Al.B.: 
Resonance factorized scattering and 
roaming trajectories. ENS-LPS-335 pp.27  (1991)}

\lref\LuZ{ Lukyanov, S. and Zamolodchikov, A.: unpublished}

\lref\saleur{Saleur, H.:
A comment on finite temperature correlations in integrable QFT.
Nucl. Phys. {\bf B567},  602-610 (2000)} 

\lref\flashk{Flaschka, H. and McLaughlin, D.: Canonically conjugate
variables for the Korteweg-de Vries equation and the
Toda Lattice with periodic boundary conditions.
Progress Theor.
Phys. {\bf 55}, 438-456 (1976)}
 
\lref\sklyan{Sklyanin, E.K.: The quantum Toda chain.
Lect. Notes in Physics {\bf 226}, 196-233 (1985)\semi
J. Sov. Math. {\bf 31}, 3417 (1985)}

\lref\feodi{Smirnov, F.A.: Structure 
of matrix elements in quantum
Toda chain. J. Phys. A.: Math. Gen. {\bf 31} 8953-8971 (1998) }
 
\lref\BLZ{Bazhanov, V.V., Lukyanov, S.L. and Zamolodchikov, A.B.:
Integrable structure of conformal field theory,
I. Quantum KdV theory and
thermodynamic Bethe Ansatz.
Commun. Math. Phys. {\bf 177}, 381-398 (1996)\semi
II. Q-operator and DDV equation. 
Commun. Math. Phys. {\bf 190}, 247-278 (1997)
\semi
III. The Yang-Baxter Relation. 
Commun. Math. Phys. {\bf 200}, 297-324 (1999)
}

\lref\BLZZ{Bazhanov, V.V., Lukyanov, S.L. and Zamolodchikov, A.B.:
Quantum field theories in finite volume: Excited state energies.
Nucl. Phys. {\bf B489} [FS],  487-531 (1997)}

\lref\pokr{Patashinskii, A.Z. and Pokrovskii, V.N.:
Fluctuation theory of phase transition,
Oxford: Pergamon Press 1979}

\lref\ZZ{Zamolodchikov, A.B. and Zamolodchikov, Al.B.:
Structure Constants and Conformal Bootstrap in
Liouville Field Theory.
Nucl. Phys. {\bf B477}, 577-605   (1996) }

\lref\LZ{Lukyanov, S. and Zamolodchikov, A.:
Exact expectation values of local fields
in quantum sine-Gordon model.
Nucl. Phys. {\bf B493}, 571-587 (1997)}

\lref\smirno{Smirnov, F.A.:  Quasi-classical 
study of form-factors in finite volume.
(Paris U., VI-VII). LPTHE-98-10, Feb 1998, pp.21 
(hep-th/9802132)  }

\lref\lecmus{Leclair, A. and Mussardo, G.:
Finite temperature correlation functions in integrable QFT.
Nucl. Phys. {\bf B552}, 624-642  (1999)}

\Title{\vbox{\baselineskip12pt\hbox{RUNHETC-2000-17}
                \hbox{hepth/nnnmmyy}}}
{\vbox{\centerline{}
\centerline{ Finite Temperature
Expectation Values}
\centerline{of Local Fields in the sinh-Gordon model} }}
\centerline{}
\centerline{Sergei Lukyanov }
\centerline{}
\centerline{Department of Physics and Astronomy,
Rutgers University}
\centerline{ Piscataway,
NJ 08855-0849, USA}
\centerline{and}
\centerline{L.D. Landau Institute for Theoretical Physics}
\centerline{Kosygina 2, Moscow, Russia}
\centerline{}
\centerline{}

\bigskip
\centerline{\bf{Abstract}}

\bigskip

Sklyanin's  method of separation of variables is  employed in
a calculation of
finite temperature
expectation values. An essential element
of the approach is  Baxter's $Q$-function.
We propose its explicit form corresponding to
the ground state of the sinh-Gordon theory.
With the  method of separation of variables we calculate
the finite temperature
expectation values  of the exponential 
fields to one-loop order of
the semi-classical  expansion.

\centerline{}

\Date{April, 00}

\eject

\newsec{Introduction}

One-point functions
have numerous applications
in Statistical Mechanics and 
Condensed Matter Physics\ \refs{\pokr,
\tzev}.
They determine ``generalized susceptibilities'' i.e. the linear
response of a system to external fields.
In a  path integral formulation  the  one-point function
of a local field ${\cal O}$ is 
represented  by a  Euclidean path
integral of the form
\eqn\saaldi{\langle\,
{\cal O}\, \rangle=Z^{-1}\
\int {\cal D}[\, \varphi\, ]  \ {\cal O}\ e^{-{\cal A}}
\ .}
Recently some progress has been
achieved in the  calculation of
one-point functions  in
integrable Quantum Field Theory (QFT)  
defined on a  two dimensional
Euclidean plane\ \refs{\LZ, \FLZZZ}. 
In this case, the integral\ \saaldi\
can also  be  viewed as  a Vacuum Expectation Value  (VEV)
in  $1+1$-dimensional QFT associated with the action ${\cal A}$.
For many applications, especially in Condensed Matter Physics,
it is important to generalize
the results of\ Refs.\refs{\LZ, \FLZZZ}
to the case of  Euclidean path integrals
defined on an infinite cylinder. In the Matsubara imaginary
time formalism such path  integrals are interpreted
as  thermal averages
\eqn\skdui{\langle\, {\cal O}\, \rangle_R=
{{\rm Tr}\big[\, e^{-R{\bf H}
}\,
{\cal O}\, \big]\over
{\rm Tr}\big[\, e^{-R{\bf H}
}\, \big]}\ ,}
where ${\bf H}$ is the Hamiltonian of the corresponding
QFT and the temperature coincides with the  
inverse circumference of
the cylinder.

The path integral defined on a cylinder
also allows  another interpretation.  It
is an expectation value for  the  ground state\
$|\, vac\, \rangle_R$\  of
the $1+1$-dimensional
theory  in the finite geometry where the spatial coordinate is
compactified on a
circle.
Hence, the  VEVs contain important
information about Renormalization
Group flow  controlled by the parameter $R$.

An exact calculation of the finite volume (finite $R$) VEVs
is a  challenge even in integrable QFT.
Recent progress made in
papers\ \refs{\smirno, \lecmus}\ should
be mentioned here.
In\ \lecmus\ A. Leclair and G. Mussardo 
proposed  an integral representation
which makes it  possible  to generate
a  low-temperature ($R\to\infty$)  expansion for the  VEVs
in terms of infinite volume form-factors of local  fields and
some thermodynamical data.
Their  conjecture works for
theories with trivial $S$-matrices  such  as Ising and
Free Dirac Fermion  models\ \LuZ, but its
validity  remains
questionable for models with non-trivial scattering amplitudes
(see e.g. \saleur).
Another line of  research was proposed in the work\ \smirno.
F. Smirnov   applied the method  
of separation of variables\ \refs{\sklyan,\sklynn}\ 
to the semi-classical  study  of
finite volume  matrix elements in the quantum KdV equation.
The  model does not constitute  a relativistic field theory.
Nevertheless,
it is of prime importance for
the sinh-Gordon QFT  since both of the  equations are
in the same integrable hierarchy.

In this paper we will implement 
the method  of separation of variables in
the case of 
the quantum  sinh-Gordon theory.
The problem  is defined  by a Euclidean action,
\eqn\sldi{{\cal A}_{{\rm shG}}=\int_{-\infty}^{\infty}dx_1
\int^{R}_{
0}dx_2\
\bigg\{\, { 1\over 16\pi}\, \big(\, \partial_{\sigma}
\varphi\, \big)^2 +2\, \mu\ \cosh(b\varphi)\
\bigg\}\  ,}
where $\varphi$ is a scalar field with  periodic boundary
condition along the $x_2$-coordinate.
We are  focusing on the VEVs of  the  exponential fields,
\eqn\lski{{\cal O}=e^{a\varphi}\ .}
For our purposes it will be useful to rewrite\ \skdui\  in the
form,
\eqn\fdddo{\langle\, e^{a\varphi}\, \rangle_R=
Z^{-1}\ \int {\cal D}[\, \chi\,  ]\ 
\Psi^2_0[\, \chi\, ]\, e^{a\chi}\ .}
Here\ $\Psi_0[\, \chi\, ]$  is
an integral taken over  field configurations on the half-cylinder,
$x_1<0$, satisfying the boundary condition,
$$\varphi(x_1, x_2)|_{x_1=0}=\chi(x_2)\ ,$$
i.e. it
is a wave functional  corresponding to the
ground state $|\, vac\, \rangle_R$.
The method  of separation of 
variables\ \refs{\sklyan,\sklynn,\smirno}\ allows one to
introduce a  change of 
integration variables in\ \fdddo,
$$\chi(x_2)\to \big\{\gamma_k\big\}_{k=-\infty}^{\infty}\, ,$$
from the function $\chi(x_2)$  to 
the infinite discrete set of $\gamma_k$.
A notable advantage of the new 
variables is that the wave functional in the
``$\gamma$-representation'' has a factorizable form,
$$\Psi_0\big[\,\{\gamma_k\}\, 
\big]\sim \prod_k {\cal Q}[\gamma_k]\ .$$
Notice that the integration
measure ${\cal D}\big[\, \{\gamma_k\}\, \big]$\ does 
not factorize in the
variables  $\gamma_k$.
At this moment, we are not able to elaborate on
all steps of the  changing of variables   on a rigorous basis.
Therefore, we suggest the deduced integral representation  for
\ $\langle\, e^{a\varphi}\, \rangle_R$ as a
conjecture rather than a
well established result.
To test the   validity of this
integral representation,
we carry out a semi-classical expansion of the VEV.

More explicitly, the parameter $b^2$  in the
action\ \sldi\ can readily be  
identified with the  Planck constant.
Then, for finite\  $\alpha=a\,  b$\ 
and $b^2\to 0$,
the  functional integral\ \saaldi\ is dominated by a
non-trivial  saddle-point configuration and admits the
semi-classical expansion,
\eqn\ksidu{\langle
\, e^{a\varphi}\, \rangle_R=
e^{-{S\over b^2}}\  D\ \big(\, 1+O(b^2)\, \big)\ .}
Here $S$\ is a
Euclidean  action on the cylinder
evaluated in the saddle-point configuration
and
the pre-exponential factor $D$\   is the result of
evaluating   the functional integral\ \saaldi\ in
the Gaussian approximation around  the classical solution.
With the proposed  integral representation we  
calculate the functions
$S$ and $D$ and find complete agreement
with the expected high- and low-temperature behavior of the VEVs.
In particular, our result 
matches well  with  the Leclair-Mussardo conjecture.

\newsec{Integral representation for VEVs}

\subsec{Flaschka-McLaughlin variables}

In the paper\ \flashk\ H. Flaschka and D. McLaughlin
found remarkable  canonically conjugate variables
in the phase spaces  of the classical Toda chain and
KdV equations.   
Their approach  can be straightforwardly adapted
to the classical sinh-Gordon equation.
Here we give a brief  review of  the
Flaschka-McLaughlin variables for 
this model. For more information and
proofs, the reader is referred to Refs.\refs{\flashk, \Fa}.

The sinh-Gordon
equation admits a zero curvature formulation: There exists
a $sl(2,{\Bbb R})$-valued connection 1-form, depending
on an auxiliary parameter $\lambda$, 
such that the condition of vanishing
curvature is equivalent to the equation of motion.
Dealing with the theory on  cylinder, one can integrate this
1-form along
some cycle, say,
\eqn\aslisui{ \ x_1=0\, , \ \ \ \ \ \ 0\leq x_2<R\, ,}
and obtain the so-called monodromy matrix,
\eqn\spdo{{\bf M}(\lambda)=
\pmatrix{{\tt A}(\lambda) & 
\lambda\, {\tt B}(\lambda) \cr \lambda^{-1}\,
{\tt C}(\lambda) & {\tt D}(\lambda)} \in SL(2,
{\Bbb R})\ \ \ \ \ \ \ (\, \Im m\, \lambda=0\, )\ .}
This matrix satisfies  important
analytical conditions which are readily obtained from
an  explicit form of the
connection.
In particular, the matrix elements in\ \spdo\ are real
analytical functions of the variable\
$\lambda^2$\
with two essential singularities at  the points
$\lambda^2=0,\ \infty$.
Zeroes of  $ {\tt B}(\lambda)$,
\eqn\assoo{\lambda^2_k\, :\ \  {\tt B}(\lambda_k)=0\ ,}
are of prime importance in the construction.
It is
possible to
show that all zeroes\ \assoo\ are simple, real, positive and
accumulate towards the essential singularities.
Thus we can order them,
$$0\leftarrow\ldots \lambda^2_{-N}<
\lambda^2_{-N+1}\ldots <\lambda^2_{0}<
\ldots \lambda^2_{N-1}<
\lambda^2_{N}\ldots \rightarrow +\infty $$
and define  two infinite sets:
\eqn\hsdtyt{\eqalign{&\big\{ 
\gamma_k \big\}_{k=-\infty}^{\infty}\, :\
\gamma_k=\log\lambda^2_k\, ,\cr &
\big\{ \pi_k \big\}_{k=-\infty}^{\infty}\, :\
\pi_k=4\, \log |{\tt A}(\lambda_k)|\  .}}
The mapping of the
canonical  Poisson data, $\varphi$
and\ $ \partial_{x_1}
\varphi\, $, to
the variables\ \hsdtyt\ is found 
to be  a canonical transformation, i.e.
\eqn\ksjdh{\big\{\, \pi_k\, ,\, 
\gamma_m\, \big\}=\delta_{k m}\, ,\ \ \ \
\big\{\, \pi_k\, ,\, \pi_m\, \big\}=
\big\{\, \gamma_k\, ,\, \gamma_m\, \big\}=0\ .}
Hence\ \hsdtyt\  are canonically conjugate  variables
in the phase space of the sinh-Gordon model.
At the same time,
we can treat  $\big\{ \gamma_k \big\}_{k=-\infty}^{\infty}$
as  coordinates
in the corresponding configuration space.

\subsec{$\gamma$-representation}

The Flaschka-McLaughlin variables proved 
to be useful in quantum theory as was
demonstrated in the seminal work\ \sklyan\ on the example
of the Toda chain equation (see also\ \refs{\feodi, \Leb}).
We refer to   a  quantization
in  these   variables as a quantization in  
$\gamma$-representation.
Recently  the
$\gamma$-representation  was employed to quantize   ``real''
KdV theory\ \smirno\foot{The monodromy matrix in the ``real''
KdV model has the form\ \spdo, whereas the  monodromy matrix of
the ``imaginary'' equation
belongs to the group $SU(2)\ (\Im m\, \lambda=0)$.
The imaginary KdV model is
related to  the sine-Gordon theory and
(perturbed) CFT with
the central charge $c<1$ (see e.g.  \BLZ).
A sensible  $\gamma$-representation 
for the imaginary equation has,
to our knowledge, not been found.}.
In fact, Smirnov presented heuristic, but
convincing,   model-independent arguments which can
also be  applied to the sinh-Gordon equation.
Following these arguments
we introduce the integral,
\eqn\siikaalsjd{
{\cal I}_N(R, a)=\prod_{k=-N}^{N}\, 
\int_{-\infty}^{+\infty}{d\gamma_{k}
\over b}
\ \prod_{N\geq k>m\geq -N}
\sinh(\gamma_k-\gamma_m)
\prod^{N }_{k=-N} \,
{\cal Q}^2[\gamma_k]\ e^{{2 \gamma_k\over b^2} ( a b+k)}\ .  }
We shall  also use slightly  different form of ${\cal I}_N$:
With the identity,
$$\prod_{N\geq k>m\geq -N}
2\,  \sinh\Big(\,
{\gamma_k-\gamma_m\over b^2}\, \Big)=
{\rm Det}\Big|\, \exp\Big(\,
 {2 k \gamma_j\over
b^2}\, \Big)\, \Big|_{-N\leq j, k\leq N}\ ,$$
the integral\ \siikaalsjd\ can be rewritten in the form
\eqn\sldoiika{\eqalign{
{\cal I}_N(R,a)&={1\over (2 N+1)!}\
\prod_{k=-N}^{N}\, \int_{-\infty}^{+\infty}{d\gamma_k\over b}
\times\cr & \prod_{N\geq k>m\geq -N}
2\, \sinh(\gamma_k-\gamma_m)\, \sinh\Big(\,
{\gamma_k-\gamma_m\over b^2}\, \Big)\
\prod^{N }_{k=-N} \,
{\cal Q}^2[\gamma_k]\ e^{{2 \gamma_k a\over b}}\ .  }}
The function 
${\cal Q}[\gamma]$ appearing in Eqs.\siikaalsjd,\ \sldoiika\
is the so-called  Baxter's $Q$-function.
It is a  non-singular function for real $\gamma$ with
leading  asymptotic behavior
\eqn\dlk{{\cal Q}[\gamma]\sim
e^{-2C_0\, \cosh(\, b q\gamma)}\, \ 
\ \ \ \ \ \ \ \ \ {\rm  as}\ \ \ \
\gamma\to \infty\ .
}
Here and below we use the notation,
$$q=b+b^{-1}\ .$$
The positive constant $C_0$ in\ \dlk\ reads explicitly,   
$$C_0=
{mR\over 4\, \sin\big({\pi b\over q}\big)}\ , $$
and $m$ is a  mass of the sinh-Gordon particle.
Therefore,\  ${\cal I}_N$\ \sldoiika\ is a 
convergent integral for
any finite $N$.
Notice that the configuration space 
of the model under consideration
is an   infinite-dimensional space.
In writing the $2N+1$-fold  integral,
we truncate it
to $2N+1$-dimensional space with  coordinates
$\big\{ \gamma_k \big\}_{k=-N}^{N}$.
As well as in the KdV theory\ \smirno,\ we
can treat
$$\Psi_N\big[\, \{\gamma_{k}\}\,
\big]=\prod^{N }_{k=-N} \,
{\cal Q}[\gamma_k]$$
as a wave functional in the
$\gamma$-representation. For $N\to \infty$,
it  corresponds 
to the ground state\
$|\, vac\, \rangle_R$ of the sinh-Gordon theory with  periodic
boundary conditions.
Furthermore, the double product in\ \sldoiika\ is an
integration measure and we shall denote it ${\cal D}_N
\big[\, \{\gamma\}\, \big]$.
It was calculated in the semi-classical approximation
in\ \smirno.
The  semi-classical analysis  suggests also that
the product
$${\cal O}_{N}=\prod^{N }_{k=-N}\, e^{{2 \gamma_k a\over b}}\ $$
represents (in the limit $N\to \infty$) 
the exponential field $e^{a\varphi}$ located at the
point $(R/2,R/2)$ on the cylinder\foot{
The position of the insertion is determined by  choosing  of the
integration contour for the monodromy 
matrix. Here  we assume that
the contour is given by\ \aslisui.}.
Therefore, the integral\ \sldoiika\ has 
the form of  a  quantum mechanical
diagonal  matrix element,
\eqn\uldii{{\cal I}_{N}=
\int  {\cal D}_N\big[\, \{\gamma\}\, \big]\
\Psi_N^{\dag}\, {\cal O}_{N}\, \Psi_N\ .}
We shall consider  the Vacuum Expectation 
Values only. In this case,
${\cal Q}$ is an eigenvalue of
the Baxter $Q$-operator corresponding to the ground state.

\subsec{ Baxter's $Q$-function in sinh-Gordon theory}

We now turn to the most delicate point of our construction:
an explicit form of the function  ${\cal Q}$.
To the best of our knowledge a rigorous  derivation of 
${\cal Q}$\ is not currently  available.
Here we formulate a conjecture for the 
sinh-Gordon $Q$-function based
on the following heuristic arguments.
First,  we note that the substitution\
$b\to  i\, \beta $\ 
transforms\ ${\cal A}_{{\rm shG}}$  \sldi\
to the action of the sine-Gordon model.
Naively  we  could  try to obtain\  ${\cal Q}$\  by means
of analytical continuation from $-1<b^2<0$.
In this coupling constant  
domain, the  Baxter $Q$-operator is relatively
well studied\ \BLZ.
Unfortunately, the sine-Gordon
$Q$-function  has an essential singularity at
$b^2=0$  the   analytical structure  
of which is  unknown.  This makes
the analytical continuation 
to the domain of the sinh-Gordon model
a highly questionable  procedure. One can  guess an explicit 
form for
${\cal Q}$ by examining
its  asymptotic behavior.
The $Q$-function in the sine-Gordon
model admits the following asymptotic expansion for $
\gamma\to+ \infty$\ \BLZ\foot{
In this work  we use a convention for the $Q$-function
which differs from the one 
of\ \BLZ\  by an  overall shift of the argument.},
$$\log Q_{sinG}\simeq-C_0\, e^{\theta}+
\sum_{n=1}^{\infty}\ C_n\ {\Bbb I}_{2 n-1}\, e^{-(2n-1)
\theta}+ \sum_{n=1}^{\infty}\ 
{\tilde {\Bbb G}}_{ n}\ e^{2n\theta (b^{-2}+1)}
\ \ \ \ \ (-1<b^2<0)\ . $$
Here we introduce a new  notation,
\eqn\salsi{\theta\equiv\gamma\ bq\ .}
The leading term of 
this expansion has already appeared in our consideration
(see\ \dlk) while
${\Bbb I}_{2 n-1}$ and $ {\tilde {\Bbb G}}_{ n}$ are
vacuum eigenvalues of the so-called
local and dual  unlocal Integrals of Motion (IM) respectively.
The constants $C_n$ depend on the normalization of 
the local IM. In Appendix A (see Eq.(A.2)) we
present their  form  for the
normalization adopted in\ \BLZ.
A similar  asymptotic form holds
for $\gamma\to- \infty$.
The eigenvalues
${\Bbb I}_{2 n-1}$  are regular functions of $b^2$,
and can be continued to the domain 
of the sinh-Gordon model without
problems.
Contrary to\ ${\Bbb I}_{2 n-1}$,
the eigenvalues of the dual  
unlocal IM,\ ${\tilde {\Bbb  G}}_n\, $,\
are highly singular functions
at $b^2=0$. One can expect that  the  appearance of these IM 
is
a consequence of the existence of   the 
soliton sector of the sine-Gordon QFT.
This sector is absent for $b^2>0$.
All these observations  suggest
the following large $\gamma$ asymptotic  behavior
in the
sinh-Gordon model,
\eqn\kdkj{\log {\cal Q} \simeq -C_0\, e^{\theta}-
\sum_{n=1}^{\infty}\ C_n\ {\Bbb I}_{2 n-1}\, e^{-(2n-1)
\theta}
\ \ \ \ \ (b^2>0)\ .}
In  Appendix A we give  numerical evidence that 
the values of local IM can be expressed in terms of a
solution of the  Thermodynamic Bethe Ansatz
(TBA) equation:
\eqn\dpdoi{C_n\ {\Bbb I}_{2n-1}=
C_0\ \delta_{n 1}+(-1)^{n}\ \int_{-\infty}^{\infty}
{d\theta\over \pi}\ e^{(2n-1)\theta}\
\log\Big(\, 1+e^{-\epsilon(\theta)}\, \Big)\ .}
Here the function  $\epsilon(\theta)$  solves the TBA equation\ 
\refs{\Zar,\Zarr,\ZZ}
\eqn\alssi{\epsilon(\theta)-mR\, 
\cosh(\theta)+\int_{-\infty}^{\infty}
{d\theta'\over 2\pi}\ \Phi(\theta-\theta')\
\log\Big(1+e^{-\epsilon(\theta')}\,
\Big)=0\ ,}
with the kernel
\eqn\dui{\Phi(\theta)={4\, 
\sin\big({\pi b\over q}\big)\, \cosh(\theta)\over
\cosh(2\theta)-\cos\big({2\pi b\over q}\big)}\ .}
The series\ \kdkj\ is an asymptotic expansion. 
In fact, it is  a divergent
geometrical series which can easily be summed up and
one can guess an explicit form of ${\cal Q}$:
\eqn\sddiuu{\log {\cal Q}(\theta)=
-{mR\over  2 \,   \sin\big({\pi b\over q}\big)}\, \cosh(\theta)+
\int_{-\infty}^{\infty}
{d\theta'\over 2\pi }\ {\log\big(1+e^{-\epsilon
(\theta')}\, \big)
\over \cosh(\theta-\theta')}\ .}
We leave an examination of  
the properties of this function  for
future publications.

One more aspect of the $\gamma$-representation 
deserves a comment.
The sinh-Gordon model manifests an 
important  non-perturbative symmetry.
The couplings $b$ and
$b^{-1}$ correspond to physically  indistinguishable  theories.
${\cal Q}$\ in\ \sddiuu\  is a 
self-dual function in a sense that
it is invariant under the substitution $b\to b^{-1}$.
Furthermore, it is easy to see  that
$${\cal I}_N\big|_{b}={\cal I}_N\big|_{b^{-1}}\ .$$
This supports our choice of the measure in\
\sldoiika. Strictly speaking,
the measure was obtained in\ \smirno\  in the  semi-classical
approximation.
The exact invariance
of the semi-classical
measure suggests its applicability for an arbitrary value of
the  coupling constant $b^2$.

\subsec{Large $N$ limit}

As was noted above, in writing the $2N+1$-fold  integral
\ \sldoiika,\ we truncate 
the  configuration space of the theory
to $2N+1$-dimensional  one. In fact, the truncation amounts to
an ultraviolet regularization with a momentum  cutoff
given by
$$\Lambda_N={2\pi N\over R}\ .$$ 
Now we let $N\to\infty$.
From\ \uldii,\ it is clear that
the ratio
\eqn\poudiu{
{\bar {\cal I}}_{N}(R,a)= 
{{\cal I}_{N}(R,a)\over {\cal I}_{N}(R,0)}\ }
represents  the VEV\ 
$\langle\, e^{a\varphi}\, \rangle_R$\ in the large $N$ limit.   
More explicitly, dimensional
analysis
suggests that
$${\bar {\cal I}}_{N}(R,a)\propto  
\Big( {\Lambda_N\over m} \Big)^{2 a^2}\ ,$$
thus  the  correct relation   has the form
\eqn\sdo{\langle\, e^{a\varphi}\, \rangle_R=\kappa_a\
\  {\rm lim}_{N\to\infty} \Big({4\pi N\over mR}\Big)^{-2 a^2}\
{\bar {\cal I}}_N(R, a)\ .}
Here $ \kappa_{a}$ is an arbitrary $R$-independent constant.
To eliminate this  ambiguity
one has
to impose some normalization condition on the fields.
For example, the so-called  
conformal normalization  stipulates that
the exponential fields with sufficiently small $|a|$ are
normalized in accordance 
with the short distance behavior of the
two-point function 
$$\langle\, e^{a\varphi}(x)\, e^{-a\varphi}(y)\,
\rangle\to |x-y|^{4 a^2}\ 
\ \ \ \ \ \ {\rm  as}\ \ \ \ |x-y|\to 0\ .$$
The key  result of\ Refs.\refs{\LZ,\FLZZZ}\  is a  calculation
of the  limit,
\eqn\oskuu{\lim_{R\to\infty}\langle\, e^{a\varphi}\, \rangle_R=
{\cal G}_a\ }
with this normalization. Explicitly,
\eqn\ksjhh{\eqalign{
{\cal G}_{a}=&
\bigg[\,
{m\,\Gamma\big(
{1\over 2b q}\big)\,
\Gamma\big(1+{b\over 2q}\big)\over 4\sqrt{\pi}}
\, \bigg]^{-2a^2}
\times\cr
&\exp\biggl\lbrace\int_{0}^{\infty}{{dt}\over t}
\bigg[\ - {{\sinh^2 ( 2ab t )}\over{2\sinh(b^2 t)\, \sinh(t)\,
\cosh(q bt)}}+
2a^2\,e^{-2t}\ \bigg]\, \biggl\rbrace\ .}}
Once we adopt the conformal normalization for the
exponential fields,
the constant $\kappa_a$  is uniquely 
determined by the  condition\ \oskuu.
In particular, the semi-classical consideration (see below) 
leads to the relation,
$$\kappa_a={\cal G}_a\ \big(\, 1+O(b^2)\, \big)\ .$$

\newsec{Semi-classical expansion} 

In this section we  will study 
the VEVs in the semi-classical approximation.
The VEV\ \sdo\  can be
represented by the Euclidean path integral\ \saaldi\  on
a  cylinder with ${\cal A}$ and ${\cal O}$ given 
by\ \sldi,\ \lski.
For fixed\ $\alpha$, 
\eqn\ksiu{\alpha=a\,  b }
and $b^2\to 0$, 
the  path integral is dominated by a saddle-point configuration
$\phi=b\, \varphi$ and
the VEV
has the form\
\ksidu,\ where $S$\ coincides with the regularized
Euclidean classical action on the cylinder:
\eqn\doign{ S=
\lim_{\varepsilon\to 0}\bigg[\,
\int_{|x-y|>\varepsilon} 
{d^2x\over 8\pi}\, \bigg\{\, {(\partial_{\sigma}\phi)^2
\over 2}
+m^2\, \sinh^2\Big({\phi\over 2}\Big)\, \bigg\}+
\alpha \oint_{|x-y|=\varepsilon} {ds\over 2\pi} \ \phi
-2\,  \alpha^2\, \log\varepsilon\, \bigg].}
Here the function $\phi$ is a solution of the classical equation
of motion,
\eqn\ytrsldi{\partial_{\sigma}^2\phi=m^2\, \sinh(\phi)\, , }
such that,
\eqn\alaias{\phi\to \, 4\alpha\ \log|x-y|
+O(1)\, \ \ \ \ 
{\rm as}\ \ |x-y|\to 0\ ,}
and
\eqn\kksuy{\eqalign{&\phi\to 0\, \ \ \ \ 
{\rm as}\ \ |x-y|\to \infty\ ,\cr
&\phi(x_1, x_2+R)=\phi(x_1, x_2)\ .}}
The field configuration\ $\phi$\ develops a singularity
at the point $y$ where the 
exponential field is inserted. Therefore,
in the definition\ \doign\ 
we cut the small disc of 
radius $\varepsilon$ around this point and
add the boundary term to the action  to ensure\ \alaias.
We also add a field independent 
term such that the action is finite
at $\varepsilon\to 0$.
The pre-exponential factor $D$\ \ksidu\  is a result of
evaluating   the path integral\ \saaldi\ in 
the Gaussian approximation around the
classical solution defined above,
\eqn\slsdi{ D=\Big(\, {m\, \varepsilon 
\over 2}\, e^{\gamma_E}\, \Big)^{-2\alpha^2}\ 
\bigg[\, {\rm Det}'\bigg({-\partial_{\sigma}^2+
m^2\, \cosh(\phi)\over
-\partial_{\sigma}^2+m^2}\bigg)\, \bigg]^{-{1\over 2}}\ , }
where $\gamma_E=0.577216\ldots$ is the Euler constant.
The first factor in 
\ \slsdi\ appears as a result of the mass    renormalization.

\subsec{Main semi-classical order}

We now proceed to the  semi-classical
calculation of the VEV $\langle\, e^{a\varphi}\, \rangle$\
using  the representation\ \sdo.
In order to apply the saddle-point 
machinery, it is  convenient to begin
with the   form\ \siikaalsjd\ for\ the integral\ ${\cal I}_N$.
To the  lowest semi-classical order,
\eqn\ldii{{\cal Q}[\gamma]\sim 
e^{-{r\cosh (\gamma)\over 2\pi b^2}
}\ . }
Here and below the notation
\eqn\sldo{r=mR }
is used.
The corresponding saddle-point equations have the form,
\eqn\slkdiui{r\, \sinh(\rho_k)=2\pi\, 
(k+\alpha)\, ,\ \  \ k=0,\pm1,\ldots,
\pm N\ .}
In writing\ \slkdiui\  we have assumed that $b^2\to 0$
and\ $\alpha=a b$\
is
fixed.
Thus we find,
\eqn\lsiodi{\log {\cal I}_N=  - {1\over b^2}\
\sum_{k=-N}^{N}\, \Big\{\, {r\over \pi}\, \cosh(\rho_k)-
2\, \rho_k\, (k+\alpha)\, \Big\}+O(1)\ .}
The sum can be evaluated with the result,
\eqn\sidui{\eqalign{
-&\sum_{k=-N}^{N}\, \Big\{\, {r\over \pi}\, \cosh(\rho_k)-
2\, \rho_k\, (k+\alpha)\, \Big\}=T-\Big({r\over 2\pi}\Big)^2\,
\log \Big({4\pi N e\over r}\Big)+2\, \alpha^2\, \log N+\cr &
\Big(\, 2N(N+1)+{1\over 3}+2\alpha^2\, \Big)
\ \log\Big({4\pi \over r}\Big)+4\, \sum_{k=1}^{N} k\, \log (k/e)
-4 \log A_G+O\Big({1\over N}\Big)\ ,}}
where $A_G=1.282427\ldots$\
is the Glaisher constant and
the function $T=T(r,\alpha)$\ reads explicitly,
\eqn\sldsdii{T=r\ 
\int_{-\infty}^{\infty}
{d\tau\over\pi^2} \ \tau\, \sinh(\tau)\
\Re e\big[\, \log\big(\, 1-e^{-r\cosh(\tau)+2\pi i\alpha}
\, \big)
 \big]\, .}
Now, combining Eqs.$\sdo,\, \lsiodi,\, \sidui$\ one obtains
$$ \langle\, 
e^{a\varphi}\, \rangle_R\sim e^{-{S\over b^2}}\ ,$$
with
\eqn\sldoi{ S=S_0(\alpha)-T(r,\alpha)+T(r,0)\ ,}
and
\eqn\sldioi{S_0=2\alpha^2\, \log\Big({m\over 4}\Big)
+\int_{0}^{\infty}{dt\over t}\ \bigg\{
\,   {{\rm sinh}^2(2\alpha t)\over t\ \sinh(2 t)}
-2 \alpha^2 e^{-2 t}\, \bigg\}\ .}
Thus the function\ \sldoi\
coincides with  the regularized Euclidean action\ \doign\foot{
This result  was  obtained by
a different method in\ \LuZ.}.

\subsec{Semi-classical expansion of {\cal Q}-function}

To compute the VEV to one-loop order, 
we have to find the
next term in the semi-classical  expansion of  ${\cal Q}$.
It can be obtained by means of 
the TBA equation\ \alssi. The kernel
$\Phi$\ \dui\ allows  an expansion  in $b^2$,
\eqn\ldih{\Phi(\theta)=
2\pi\delta(\theta)+2\pi b^2\ \ 
P.V.\, {\cosh(\theta)\over\sinh^2(\theta)}+
O(b^4)\ , }
thus to lowest  order the solution of the TBA equation
has the form,
$$e^{\epsilon}=e^{r\cosh(\theta)}-1 +O(b^2)\ .$$
With this equation and the definition\ \sddiuu\
we calculate
\eqn\sddiu{\log {\cal Q}[\gamma]=
-{r \cosh(\gamma)\over  2   \pi\, b^2}-
{r \cosh(\gamma)\over  2 \,   \pi }+
{r \gamma\sinh(\gamma)\over  2 \,   \pi }-
\int_{-\infty}^{\infty}{d\tau\over 2\pi }\ 
{\log\big(1-e^{-r \cosh(\tau)
} \big)
\over \cosh(\gamma-\tau)}+O(b^2)  
\,  .} 
In order to see an analytical structure
of this function it is instructive 
to represent\ ${\cal Q}$\ in the form of an infinite
product,
\eqn\kksldoi{\eqalign{{1\over  {\cal Q}[\gamma]}=
&e^{{r \cosh(\gamma)\over  2 \pi b^2}}\
\Big({r e^{\gamma_E}\over 4\pi}\Big)^{ {r \cosh(\gamma)
\over 2\pi}}\ \sqrt{2r}\times \cr &
\cosh\Big({\gamma\over 2}\Big)\
\prod_{n=1}^{\infty}\, 
\bigg\{\, \sqrt{1+\Big({r\over 2\pi n}\Big)^2 }
+{r\cosh(\gamma)\over 2\pi n}\, \bigg\}
e^{-{r\cosh(\gamma)
\over 2\pi n}}\ \ \big(\, 1+O(b^2)\, \big)\ .}}
Here $\gamma_E$ is the Euler constant.

\subsec{ One-loop order}

The saddle-point approximation 
allows one  to  find the one-loop 
order in the semi-classical expansion.
To this 
order,
\eqn\spdo{{\cal I}_N= W\   
\prod_{N\geq k>m\geq -N}
\sinh(\rho_k-\rho_m)
\ \prod^{N }_{k=-N} \,
{\cal Q}^2[\rho_k]\ 
e^{2 \rho_k (\alpha+k) }\ \Big(\, 1+O(b^2)\, \Big)\ .}
Here the function $W$ is a result of the Gaussian
integrations in\ \siikaalsjd\ around 
the saddle points $\gamma_k=\rho_k$\ \slkdiui,
$$W=(2\pi^2)^{N+{1\over 2}}
\ \prod_{k=-N}^{N} {1\over \sqrt{ r\cosh(\rho_k)}}\  .$$
The main steps in the
calculation of\ \spdo\ are given in
Appendix B.
Our final result has the form \ \ksidu\
with  the function $ S$  given by\ \sldoi\ and
\eqn\sdiui{\eqalign{\log 
D=&\log D_0+T(r,\alpha)-T(r, 0) -\alpha\,
\partial_{\alpha}T(r,\alpha)
-\cr &
\int_r^{\infty}{dr\over 8}\ r\, \bigg\{\,
\big(\,\partial_r\partial_{\alpha}T\,\big)^2
-{1\over 2\pi^2}\ \Big(\, 
\partial_{\alpha}^2T\big|_{
\alpha=0}\   \partial_{\alpha}^2T-
\big(\,\partial^2_{\alpha}T\,\big)^2 \big|_{\alpha=0}\, \Big)
\, \bigg\}\ .}}
Here
$$\log D_0=- 2 \alpha^2\, \log 2+
{1\over 2}\ \int_{0}^{\infty}dt\ {\sinh^2
(2 \alpha t)
\over \cosh^2(t)}\ .$$
Recall that\ \sdiui\ should coincide 
with  the functional determinant \slsdi.

\newsec{High-temperature behavior}

Now that we have computed\ \ksidu,
let us check the result for some limiting cases.
Here we argue  for   $R\to 0$ behavior of the VEVs.

Due to the scaling properties 
of the interaction operator in\ \sldi\
one can rescale the problem to a circle of circumference $2\pi$.
Thus, the Hamiltonian of 
the model under consideration takes the form
\eqn\saaldi{{\bf H}_{{\rm shG}}={2\pi\over R}\
\int^{2\pi}_{
0}d\tau\
\bigg\{\, 4\pi \, \Pi^2
 + { 1\over 16\pi}\,  \big(\partial_{\tau}
\chi)^2 + \mu\ \Big({R\over 2\pi}\Big)^{2bq}\ 
\big( \, e^{b\chi}+
e^{-b\chi}\, \big)\ \bigg\}\ ,}
where $\Pi={1\over i}\, {\delta\over\delta\chi}$ is 
the  momentum conjugate to
$\chi=\varphi|_{x_1=0}$.
The mass of the sinh-Gordon particle 
is related to the parameter $\mu$ by
\Zarn,
\eqn\uyodi{\mu=-{\Gamma(-b^2)\over \pi\Gamma(1+b^2)}\
\bigg[{m\, \Gamma(
{1\over 2bq})\Gamma(1+{b\over 2q})
\over 4\sqrt{\pi}}\bigg]^{2bq}\ .}
For $r\to 0$ and $a>0$ the main 
contribution in the path integral
\fdddo\  comes from a region of the 
configuration space corresponding to
\eqn\skdju{\chi\sim -2 q\  \log(r) \gg 1\ .}
In this region we can neglect the term  $e^{-b\chi}$ in the
Hamiltonian\ \saaldi, and approximate  
the ground state wave functional
$\Psi_0[\chi]$ by a proper wave 
functional from the Liouville Conformal
Field Theory (CFT).

More explicitly, the Hilbert space of the Liouville CFT
contains a continuous  set of
primary states
$|\, p\, \rangle$ parameterized by
$p>0$ with the conformal dimension\ \cortt,
\eqn\ksdhy{\Delta_p=p^2+{q^2\over 4}\  .}
We will assume that these states  are canonically  normalized,
$$\langle \, p'\, |\, p\, \rangle=2\pi\ \delta(p-p')\ .$$
Let $\Psi_p[\chi]$ be a normalized 
wave functional corresponding to the
state $|\, p\, \rangle.$
As was discussed in
\ \refs{\Zarr,\ZZ},\ the following relation holds
\eqn\jdhfy{\Psi_0[\chi]\approx L_p\ \Psi_p[\chi]\ }
in the region\ \skdju.
Here $p=p(r)$ solves  the  equation,
\eqn\sldii{p(r)\, :\,
2pq\, \log\bigg[\, 
{r\Gamma({1\over 2 bq})\Gamma(1+{1\over 2 bq})\over
8\pi^{{3\over 2}}\  
(\, b^2\, )^{{1\over  bq}}}\, \bigg]=-{\pi\over 2}+
\Im m\, \Big[ \log
\big\{\, \Gamma(1+2 ip/b)\, 
\Gamma(1+2 ip b)\, \big\}\, \Big].}
We emphasize  that $L_p$ is a unique coefficient provided
a  normalization of $\Psi_0$ is chosen.
Therefore, one can  
expect the following  relation for $r\ll 1$:
\eqn\skduua{\langle\, e^{a\varphi}\, \rangle_R\approx
 L_p L_{-p}\
{\langle \, p\, |\, 
e^{a\varphi}\, |\, p\, \rangle_{{\rm Liouv}}\over
{}_R\langle\, vac\, |\, vac\, \rangle_{R} }\ .}
The   matrix element\
$\langle \, p\, |\, e^{a\varphi}\, |\, p\, \rangle_{{\rm Liouv}}$\
was found in\ \refs{ \dorn, \ZZ}. It reads explicitly,
\eqn\disksu{\langle \, p\, |
\, e^{a\varphi}\, |\, p\, \rangle_{{\rm Liouv}}=\Big({R\over 2\pi}\Big
)^{2 a (q-a)}\ 
\bigg[\, {\pi\mu\Gamma(b^2)\,
b^{2-2 b^2}\over
\Gamma(1-b^2)}\, \bigg]^{-a/b}\
{\Upsilon_0\ \Upsilon(2 a)\, 
\Upsilon(2i p )\, \Upsilon(-2 i p)\over
\Upsilon^2(a)\Upsilon(a+i p)
\Upsilon(a-i p)}\ .}
Here  we use the notations\ \ZZ
$$\log\Upsilon(a)=-\Big(\, {q\over 2}-a\, \Big)^2\
\log(2 b)+\int_0^{\infty}{dt\over t}\ \bigg[
\, \Big(\, {q\over 2}-a\, \Big)^2\,
e^{-2t}-{\sinh^2\big(\, (qb-2ab) t
\,\big)\over \sinh(2t) \sinh(2t b^2)}\, \bigg]\ ,$$
and
$$\Upsilon_0=\partial_a \Upsilon(a)\big|_{a=0}\ .$$

It is easy to see that the function $p=p(r)$
\sldii\ satisfies the condition,
\eqn\kdid{\lim_{r\to 0}\,  p(r)=0\ .}
Using Eqs.\skduua-\kdid, we can derive
\eqn\skduaaa{
\langle\, e^{a\varphi}\, \rangle_R\approx {\cal N}^2\
\bigg[{m\, \Gamma(
{1\over 2bq})\Gamma(1+{b\over 2q})
\over 4\sqrt{\pi}}\bigg]^{-2aq}
\, b^{2aq}\
\Big({R\over 2\pi}\Big)^{2a(a-q)}\
\, {\Upsilon(2 a) \Upsilon^3_0\over \Upsilon^4(a)}\ ,}
where
\eqn\utldii{{\cal N}^2=
{\lim_{p\to 0} \big\{\, 4p^2\, L_pL_{-p}\, \big\}\over
{}_R\langle\, vac\, |\, vac\, \rangle_{R} } }
does not depend on $a$.
In writing\ \skduaaa\ we also  used the
relation\ \uyodi.
Unfortunately,  the function ${\cal N}={\cal N}(r, b)$ is not
known in  closed form
for an arbitrary
value of the coupling constant.
One can  obtain its limiting value as $b^2\to 0$.
For  $ b^2\ll r\ll 1 $,
it is sufficient to consider the dynamics
of the zero-mode $X$\ \refs{ \Zarr, \ZZ}:
\eqn\ksdu{X=\int_{0}^{2\pi} {d \tau\over 2\pi}\
\chi(\tau)\ .}
In this approximation, known as the mini-superspace
approach\ \ziy, the
Hamiltonian\ \saaldi\ is substituted by,
$${\bf H}_{ms}={2\pi\over R}\ \bigg\{- 2\, \partial^2_{X}+
\Big({r\over 4\pi b}\Big)^{2}\, \cosh(b X)\, \bigg\}\ .$$
The  Schr$\ddot {\rm o}$dinger equation
$${\bf H}_{ms}\ \Psi_0(X)=E_{ms}\ \Psi_0(X)\ ,$$
coincides with the modified Mathieu equation
and the  wave functional
$\Psi_0$  is represented by
its  lowest eigenfunction.
We will use the common normalization condition
\eqn\sldoi{\int_{-\infty}^{\infty}dX\,
\Psi^2_0(X)=1\ .}
The Liouville  wave functionals\ ${\Psi}_p$\ in 
the mini-superspace
approximation have the form,
$${\Psi}_p(X)={2\over\Gamma(2ip/b)}\
\Big({r\over 8\pi b^2}\Big)^{2ip/ b}\
K_{{2ip\over b}}
\Big({r\over 4\pi b^2}\,  e^{{bX\over 2}} \Big)\ .$$
Here $K_{\nu}(z)$ is the MacDonald function.
Now it is clear that the mini-superspace approximation for
${\cal N}$\ \utldii\ can be obtained
from the large $X$ behavior of the normalized Mathieu function
\ $\Psi_0(X)$\ \sldoi,
\eqn\hyldii{\Psi_0(X)\to \sqrt{2\over r}\
{\pi\ \ {\cal N}_{ms} \over \cosh(bX/4)}\   \exp\Big\{\,
-{r\over 2\pi b^2}\,  \cosh(bX/ 2)\,
\Big\}\, \ \ \ \ \ \ {\rm as}\ \ X\to \pm \infty\  .}

Of concern to us is the behavior of ${\cal N}_{ms}$
in the domain $b^2\ll r\ll 1$. In this case,
we  replace $\Psi_0(X)$ by its WKB asymptotic\ \hyldii\ and
readily obtain,
\eqn\lksiui{\lim_{b^2\to 0}{\cal N}^2=
{\sqrt{2}\,  r^{3\over 2}\over (2\pi)^3}\ ,
\ \ \ \ \  \   r\ll 1\ .}

Having arrived at Eq.\ \lksiui,
we can  straightforwardly expand \skduaaa\ in 
a power series of $b^2$,
\eqn\ueyrt{\langle\, e^{a\varphi}\, \rangle_{R}\approx  F\
 \Big(\, 1+O(b^2)\, \Big)\ \ \ \ \ \ \ (b^2\ll r\ll1)
\ ,}
with
$$\eqalign{F=&
\Big({R\over 2\pi}\Big)^{ {2\alpha (\alpha-1)\over b^2} }\,
 2^{-{2\alpha\over b^2}}\,
\exp\Big\{ {1\over 2b^2} \, \big(
4\, S_0(1/2-\alpha)- S_0(1/2-2\alpha)\big)\Big\}
\times\cr &
2^{2+{1\over 2b^2}}\ m^{-{3\over 4b^2}}\
\ A_G^{-{9\over b^2}}\  \Big(
{r\over 4\pi}\Big)^{{3\over 2}-2\alpha}\
\sqrt{{\Gamma(1-2\alpha)\over
\Gamma(2\alpha)}}\ {\Gamma^2(\alpha)\over \Gamma^2
(1-\alpha)}\ ,}$$
where $S_0$ is given by  Eq.\sldioi\ and
$A_G$
is the Glaisher constant.
It is possible to show that the  high-temperature
expansion of\ \ksidu\  exactly 
matches\ \ueyrt\ (see Appendix B for some
details).

\newsec{Low-temperature expansion}

We have  mentioned in the Introduction 
that the finite volume  VEVs 
can be understood  as  thermal averages\ \skdui.
Hence, $\langle\, e^{a\varphi}\, \rangle_R$ admits the
low-temperature ($R\to\infty$) expansion in the form,
\eqn\dslsi{
\log\big(\,  \langle\, e^{a\varphi}\, \rangle_R/
{\cal G}_a\, \big)=
1+\sum_{k=1}^{\infty} G_k(r)\ .}
Here $G_k$ represents
$k$-particle contributions in the infinite-volume channel and
$$G_k(r)\sim e^{-kr}\ .$$
Recently A. Leclair and G. Mussardo\ \lecmus\ proposed 
an integral representation
which is sufficient 
to generate $G_k(r)$  systematically  in terms
of form-factors of the field $e^{a\varphi}$  at $R=\infty$
and the solution of the TBA equation
\alssi. 
Taking into account   contributions of
one- and two-particle states to the 
trace\ \skdui,\  they obtained
\eqn\pldii{\eqalign{\log\big(\, 
\langle\, e^{a\varphi}\, \rangle_R/
{\cal G}_a\, \big)
=&\,  4\, [a]
\ \int_{-\infty}^{\infty} {d\theta\over 2\pi}
\ f_-(\theta)+\cr &
[2\, a]\ 
\int_{-\infty}^{\infty} 
{d\theta_1\over 2\pi}{d\theta_2\over 2\pi}\ 
{\Phi(\theta_1-\theta_2)\over \cosh(\theta_1-\theta_2)}\ 
f_-(\theta_1)f_-(\theta_2)+\ldots\ ,}}
where   the notations
$$f_-(\theta)={1\over 1+e^{\epsilon(\theta)}}\ ,$$
and
$$[a]={\sin^2({\pi a\over q})\over
\sin({\pi b\over q})}\ $$
are used.
The function $\Phi$  is the kernel in the TBA equation\ \alssi.
With\ \pldii\ and the TBA equation one can
calculate  the first 
two terms in the low-temperature expansion\
\dslsi:
\eqn\skduj{\eqalign{&G_1={4\, [a]\over \pi}
\ K_0(r)\ ,\cr
&G_2=4\, [a]\ 
\int_{-\infty}^{\infty}
{d\theta_1\over 2\pi}
{d\theta_2\over 2\pi}\, \Big(\, \Phi(\theta_1-\theta_2)-
2\pi\delta(\theta_1-\theta_2)\, \Big)\, e^{-r\cosh\theta_1+
r\cosh\theta_2}+\cr&
\ \ \ \ \  \ \ \ \ \ \ \ \ \ \ \ [2\, a]\ 
\int_{-\infty}^{\infty} 
{d\theta_1\over 2\pi}{d\theta_2\over 2\pi}\
{\Phi(\theta_1-\theta_2)\over \cosh(\theta_1-\theta_2)}\ 
e^{-r \cosh\theta_1+
r\cosh\theta_2}\ ,}}
where $K_n(r)$ is the MacDonald function.
We now  expand\ \skduj\ as a  power series in $b^2$, 
\eqn\sldaai{G_1=4\, 
\bigg\{\,  {s^2(\alpha)\over b^2}+   s^2(\alpha)
-\alpha\,  s(2\alpha)
+O(b^2)\, \bigg\}\, K_0(r)\ ,}
where
$$s(\alpha)={\sin(\pi\alpha)\over \pi}\ .$$ 
To expand $G_2$ one needs to use Eq.\ldih,
\eqn\sso{\eqalign{
G_2 \, &=\bigg\{\, { s^2(2\alpha)\over b^2}+
s^2(2\alpha)-2\alpha\, s(4\alpha) 
\, \bigg\}\, K_0(2r)-
\cr &
4\,  s^2(\alpha)\ r^2\ \big(\, K^2_1(r)-K^2_0(r)\, \big)-
 s^2(2\alpha)\ r^2\ 
\big(\, K_2(r) K_0(r)-K^2_1(r)\, \big)+O(b^2)\ .}}
It is quite straightforward to verify that
the low-temperature expansion of\ \ksidu\ exactly reproduces
\ \sldaai\ and\ \sso.

\newsec{ Conclusion}

The proposed $\gamma$-representation\ \sdo\
is the  main result of this paper.  Its rigorous derivation 
has not yet been achieved.
Although\ \sdo\ are  conjectures,
the evidence presented  in this  paper appears to make it
reasonable to take them as the starting point
for further investigation.

One can  expect that similar representations
exist for non-minimal CFT, 
say,   the Liouville theory and $SL(2,
{\Bbb R})/U(1)$
non-compact $\sigma$-model.
It may  cast new light  on many unsolved
problems of  $2D$ Quantum Gravity.
In this connection an intriguing similarity between
the integrals appeared in the $\gamma$-representation and
Matrix Models of  $2D$ Quantum 
Gravity\ \refs{\Briz,\Dug,\Gros}\ can be mentioned.

\vskip0.3cm

\centerline{\bf Acknowledgments}

\vskip0.3cm

I am grateful to A.B. Zamolodchikov for 
interesting discussions.
The research  is supported 
in part by the DOE grant \#DE-FG05-90ER40559.

\bigskip

\centerline{{\bf Note added}}

\vskip0.3cm

After finishing this paper it was drawn 
to my attention that Al.B. Zamolodchikov   independently
introduced
and studied the function\ \sddiuu\ in Ref.\zama.
I am grateful to him  for the  communication of that paper, and
sharing insights.

\bigskip
 
\appendix{A}{}

The QFT defined by\ \sldi\  possesses 
infinitely many local IM\ ${\hat {\Bbb I}}_{2
n-1}$
whose vacuum eigenvalues  have appeared in the equation\ \kdkj.
They can be represented in the form,
$${\hat {\Bbb I}}_{2 n-1}=\int_0^{R} {dx_2\over 2\pi}\ \Big(\,
T_{2n}(x_2+ix_1, x_2-ix_1)+
\Theta_{2n-2}(x_2+ix_1, x_2-ix_1)\, \Big)\ ,$$
where the local 
fields $T_{2n}$ and $\Theta_{2n-2}$ satisfy the continuity
equations,
$$\partial_{{\bar z}}T_{2n}(z,{\bar z})=\partial_{ z}
\Theta_{2n-2}(z,{\bar z})\ .$$
Although a general expression for 
the densities\ $T_{2n}$,  $\Theta_{2n-2}$\ is
not known, they are determined  up to  normalization by the
commutativity conditions,
$$\big[\, {\hat {\Bbb I}}_{2 n-1}\, ,
\, {\hat {\Bbb I}}_{2 m-1}\, \big]=0\ .$$
In Refs.\refs{\BLZ,\BLZZ}\ the 
following normalization was adopted,
\eqn\ski{T_{2n}=2^{-2n}\ (\partial_z\varphi)^{2n}+\ldots\ ,}
where  omitted terms contain higher derivatives
of $\varphi$ and exponential
fields. Notice that the condition\ \ski\ does not depend on the
regularization
scheme defining
the composite field $(\partial_z\varphi)^{2n}$.
With  normalization\ \ski\ the 
constants $C_n$\  \kdkj\  were found in
Refs.\refs{\Al,\BLZ},
\eqn\ildoi{C_n={
\Gamma\big(\, {(2n-1)\,  b\over 2q}\, \big) \,
\Gamma\big(\, {2n-1 \over 2bq }\, 
\big)\over 2\ \sqrt{\pi}\  n!\  q}\
\biggl[\,{m\ \Gamma\big({ b\over 2q}\big) \,
\Gamma\big({1 \over 2q b}\big)\over 8\, q\,
\sqrt{\pi} }\, \biggr]^{1-2 n}\ .}

The local IM  ${\hat {\Bbb I}}_{2 n-1}$ are certain
deformations of the local IM  of the Liouville CFT.
Let $I_{2 n-1}(p)$ be an  
eigenvalue of the Liouville local IM corresponding
to the state $|\, p\, \rangle$\ \ksdhy, 
while ${\Bbb I}_{2 n-1}$ is   the
sinh-Gordon ground state 
eigenvalues of ${\hat {\Bbb I}}_{2 n-1}$. 
Eq.\jdhfy\ suggests the following relation for $r\ll 1$:
\eqn\ldiik{{\Bbb I}_{2 n-1}=
I_{2n-1}\big(\, p(r)\, \big)+O\big(r^{4qb}, r^{4q/b}\big)\ ,}
where $p(r)$  solves\ \sldii\foot{For $n=1$
this relation was discussed in Ref.\ZZ.}. The
power corrections in $r$\ \ldiik\ can be
obtained  by means of Conformal Perturbation Theory.
Explicit forms of the 
functions $I_{2 n-1}(p)$ (for $n=1,\ldots ,8$)
are given
in  Appendix B of Ref.{\BLZZ}. 
Here we present only the first two of them,
$$\eqalign{&I_{1}(p)={2\pi \over R}\
 \Big(\, p^2-{1\over 24}\, \Big)\ ,\cr
&I_{3}(p)=\Big(\, {2\pi \over R}\, 
\Big)^3\ \Big(\, p^4-{p^2\over 4}+
{4 b^4+17 b^2+4\over 960\, b^2}\, \Big)\ .}$$

In  Tables 1-4 we list  numerical values of
the local IM  ${\Bbb I}_{2 n-1}$
for some $0.01\leq r\leq 1$
and $b^2=0.81$ which were obtained 
by means of numerical solution of the
TBA equation\ \alssi\ with  use of Eqs.\dpdoi\   and\ \ildoi.
These  data are compared against   values of the
Liouville local IM $I_{2n-1}\big(\, p(r)\, \big)$.
We consider the content of  
Tables 1-4 to be  an impressive evidence in support of
the relation\ \dpdoi.

\vfil

\eject

\midinsert
\centerline{
\noindent\vbox{\offinterlineskip
\def\tablerule{\noalign{\hrule}}
\halign{
\strut#&\vrule#\tabskip=1em plus2em&
   #&\vrule#&
   #&\vrule#&
   #&\vrule#
\tabskip=0pt
\cr\tablerule
&& $r$
&&\ ${\Bbb I}_1 $
&&\ $I_1\big((p(r)\big)$ &
\cr\tablerule
&& 1.0   && -0.0059897196942029 && -0.0059891933248581 &
\cr\tablerule
&& 0.8   && -0.0123637695005731 && -0.0123636397906571 &
\cr\tablerule
&& 0.6   && -0.0183955019094376 && -0.0183954816200409 &
\cr\tablerule
&& 0.4   && -0.0242191017573388 && -0.0242191003738075 &
\cr\tablerule
&& 0.2   && -0.0301582132435655  && -0.0301582132312211 &
\cr\tablerule
&& 0.1   && -0.0335250692109914  && -0.0335250692108919 &
\cr\tablerule
&& 0.01  && -0.0381898149656469  && -0.0381898149656469 &
\cr\tablerule
\noalign{\smallskip} }
}
}
\vskip 25pt
\centerline{
\noindent\vbox{\offinterlineskip
\def\tablerule{\noalign{\hrule}}
\halign{
\strut#&\vrule#\tabskip=1em plus2em&
   #&\vrule#&
   #&\vrule#&
   #&\vrule#
\tabskip=0pt
\cr\tablerule
&& $r$
&&\ ${\Bbb I}_3 $
&&\ $I_3\big((p(r)\big)$ &
\cr\tablerule
&& 1.0   && 0.01857760504756  && 0.01858088002375 &
\cr\tablerule
&& 0.8   && 0.01975951264163  && 0.01976027692024  &
\cr\tablerule
&& 0.6   && 0.02095100471765  && 0.02095111804696 &
\cr\tablerule
&& 0.4   && 0.02216988492643  && 0.02216989225147 &
\cr\tablerule
&& 0.2   && 0.02348269733500    && 0.02348269739676 &
\cr\tablerule
&& 0.1   && 0.02425825249985  && 0.02425825250033 &
\cr\tablerule
\noalign{\smallskip} }
}
}
\vskip 25pt
\centerline{
\noindent\vbox{\offinterlineskip
\def\tablerule{\noalign{\hrule}}
\halign{
\strut#&\vrule#\tabskip=1em plus2em&
   #&\vrule#&
   #&\vrule#&
   #&\vrule$\,$\vrule#&
   #&\vrule#&
   #&\vrule#
\tabskip=0pt
\cr\tablerule
&& $r$
&&\ ${\Bbb I}_5 $
&&\ $I_5\big((p(r)\big)$
&&\ ${\Bbb I}_7 $
&&\ $I_7\big((p(r)\big)$ &
\cr\tablerule
&& 1.0   && -0.02830178173  && -0.02830149724 &
& 0.0731035360   && 0.0731032652 &
\cr\tablerule
&& 0.8   && -0.02936366562  && -0.02936359281  &
& 0.0750097817  &&  0.0750097140  &
\cr\tablerule
&& 0.6   && -0.03040878256  && -0.03040877074 &
& 0.0768644292  && 0.0768644184 &
\cr\tablerule
&& 0.4   && -0.03145618592  && -0.03145618508 &
& 0.0787032289  && 0.0787032281 &
\cr\tablerule
&& 0.2   && -0.03256438579    && -0.03256438578 &
& 0.0806285556  && 0.0806285556 &
\cr\tablerule
\noalign{\smallskip} }
}
}
\vskip 25pt

\centerline{
\noindent\vbox{\offinterlineskip
\def\tablerule{\noalign{\hrule}}
\halign{
\strut#&\vrule#\tabskip=1em plus2em&
   #&\vrule#&
   #&\vrule#&
   #&\vrule$\,$\vrule#&
   #&\vrule#&
   #&\vrule#
\tabskip=0pt
\cr\tablerule
&& $r$
&&\ ${\Bbb I}_9 $
&&\ $I_9\big((p(r)\big)$
&&\ ${\Bbb I}_{11} $
&&\ $I_{11}\big((p(r)\big)$ &
\cr\tablerule
&& 1.0   &&  -0.29532264   &&  -0.29532199 &
&  1.73235   &&  1.73235 &
\cr\tablerule
&& 0.8   &&  -0.30116225  &&  -0.30116209  &
&  1.75983  &&  1.75983  &
\cr\tablerule
&& 0.6   &&  -0.30680503  &&  -0.30680500 &
&  1.78627  &&  1.78627 &
\cr\tablerule
&& 0.4   &&  -0.31236318  &&  -0.31236317 &
&  1.81220  &&  1.81220 &
\cr\tablerule
&& 0.2   &&  -0.31814537  &&  -0.31814537 &
&  1.83907  &&  1.83907  &
\cr\tablerule
\noalign{\smallskip} }
}
}
\vskip 25pt
\capt{Tables 1-4. Comparison of the  
LHS and RHS of equation\ \ldiik\ ($b^2=0.81$).  }

\vfill

\eject

\endinsert

\appendix{B}{}

Here we proceed with 
calculation of the products in\ \spdo\  and
give some technical 
hints on the  study of their high-temperature behavior.
First, let us consider the product,
$$e^{M_1}= W\ \prod_{N\geq k>j\geq-N}\sinh(\rho_k-\rho_m) \ .$$
Using the relation
$$\partial_{r}\rho_k=-{1\over r}\ \tanh(\rho_k)\ ,$$
which follows from the saddle-point equations\ \slkdiui, one
obtains
$$\partial_r M_1=
-{1\over 2r}\, \sum_{k,m=-N}^N {\cosh(\rho_k-\rho_m)
\over \cosh(\rho_k)\cosh(\rho_m) }\ .$$
This sum can be rewritten in the form,
\eqn\skddy{\partial_r M_1=
-{(2 N+1)^2\over 2r}+
{1\over 2r}\ \bigg[\,
\sum_{k=-N}^{N}\tanh(\rho_k)
\bigg]^2\ .}
Notice that the  sum in\ \skddy\ converges for $N\to\infty$,
$$\lim_{N\to \infty}\sum_{k=-N}^{N}\tanh(\rho_k)=
-{r\over 2} \partial_r\partial_{\alpha}
T+2\alpha\ .$$
To derive this formula we used
\sidui\ and the saddle-point equations\ \slkdiui.
Thus  we obtain,
\eqn\sdoo{\eqalign{M_1=&{\cal M}_N -{(2 N+1)^2\over 2}\
\log\Big({r\over 4\pi}\Big)-\cr &
2\alpha^2\ \log\Big({4\pi N\over r}\Big)-
\alpha\,
\partial_{\alpha}T(r,\alpha)-
\int_r^{\infty}{dr\over 8}\ r\,
\big(\,\partial_r\partial_{\alpha}T\,\big)^2
+O\big( N^{-1}\big)\ .}}
The constant ${\cal M}_N$ here does not depend on $r$.
To find how it depends on $\alpha$,  let us consider
$\partial_{\alpha} M_1$. A similar  calculation leads to the
equation,
\eqn\diui{\eqalign{&\partial_{\alpha} M_1={1\over 4}\,
\big(\, r\partial_r\partial_{\alpha} T-4\alpha\, \big)\,
\big(\, \partial_{\alpha}^2T+
4\, \log(4\pi N/r)\, \big)+
{2\pi\over r}\, \sum_{ m=-N}^{N} {\tanh(\rho_0)
\over
\cosh(\rho_0)+\cosh(\rho_m)}+\cr &
{2\pi\over r}\, \sum_{ m=-N}^{N}\sum_{ k=1}^{N}
\, \bigg[\, {
\tanh\big(\rho_k(\alpha)\big)
\over
\cosh(\rho_k\big(\alpha)\big)+\cosh(\rho_m) }-
{\tanh\big(\rho_k(-\alpha)\big)
\over
\cosh(\rho_k\big(-\alpha)\big)+\cosh(\rho_m) }\, \bigg]\ .}}
It follows  immediately from the last equation that
$$\partial_{\alpha}M_1\big|_{r\to \infty} =-4\, \alpha\
\log\Big( {4\pi N\over r}\Big)\ .$$
Therefore, we conclude that the constant ${\cal M}_N$ in \sdoo\
does not depend on $\alpha$.
Notice that  Eq.\diui\ is very convenient for  studying the
high-temperature limit $r\to 0$. It is
straightforward to show that for $\alpha>0$,
$$\eqalign{&\ \ \ \ \ \ \ \ \
e^{M_1}\big|_{r\to 0}\to\Big({4\pi\over r}\Big)^{2N(N+1)}\
N^{-2\alpha^2}\ e^{{\cal M}_N}\times\cr &
{2^{-2\alpha}
\over \Gamma({1\over 2}+
\alpha)}\ \sqrt{{2\pi\, \Gamma(\alpha)\over \Gamma(1-\alpha)}}\
e^{-2\alpha^2\gamma}\
\prod_{k=1}^{\infty}
{\Gamma^2\big({k+1\over 2}\big)\, e^{{2\alpha^2\over k}}
\over \Gamma\big({k+1\over 2}-\alpha\big)\,
\Gamma\big({k+1\over 2}+\alpha\big)}\ .}$$

To finish the calculation of\ \spdo\ one 
needs to evaluate the product,
\eqn\shdy{ e^{M_2}=\prod^{N }_{k=-N} \,
{\cal Q}^2[\rho_k]\ e^{2 \rho_k (\alpha+k) }\ .}
The first
two terms of the semi-classical expansion for ${\cal Q}$
are given by\ \sddiu.
With the saddle-point equation\ \slkdiui\
$M_2$\ in\ \shdy\ can be written as,
\eqn\sgdr{M_2=M'_2+M_2''\ , }
where
\eqn\sgdrf{M_2'=-\big(\,  b^{-2}+1\, \big)\
\sum_{k=-N}^{N}\, \Big\{\, {r\over \pi}\, \cosh(\rho_k)-
2\, \rho_k\, (k+\alpha)\, \Big\}   }
and
\eqn\sjdy{M_2''=-2\
\int_{-\infty}^{\infty} 
{d\tau\over 2\pi }\,  \log\big(1-e^{-r\cosh
(\tau)}\, \big)\
\sum_{k=-N}^{N}
{1\over \cosh(\rho_k-\tau)}\ .}
The sum $M_2'$ is evaluated by means of  Eq.\ \sidui.
To calculate $M_2''$ we note that
$$\partial_r M_2''=
-2\, \sum_{k=-N}^{N}\, {1\over \cosh(\theta_k)}
\
\int_{-\infty}^{\infty} {d\tau\over 2\pi }\  {m\over e^{r\cosh
(\tau)}-1}\ .$$
With the relations
$$\sum_{k=-N}^{N}\, {1\over \cosh(\theta_k)}=
{r\over 2\pi}\ \sum_{k=-N}^{N} \partial_{\alpha}\theta_k=
{r\over 4\pi}\, \bigg\{ \,
\partial_{\alpha}^2T+
4\log\Big({4\pi  N\over r}\Big)\, \bigg\}+O\Big(
{1\over N}\Big) $$
and
$$\int_{-\infty}^{\infty}
{d\tau\over e^{r\cosh
(\tau)}-1}={1\over 4}\  \partial_{\alpha}^2T\big|_{
\alpha=0}\ ,
$$
one obtains,
\eqn\uusldi{M''_2=\int_r^{\infty} {dr\over 16\pi^2}\, r\
\partial_{\alpha}^2T|_{
\alpha=0}\ \partial_{\alpha}^2T+\int_r^{\infty}
{dr\over 4\pi^2}\, r\
\partial_{\alpha}^2T|_{
\alpha=0}\ \log\Big({4\pi N\over r}\Big)\ .}
We specify the integration constant here using the condition,
$$M''_2|_{r\to\infty}\to 0\ ,$$
which follows   from the definition\ \sjdy.
The function $T$\ \sldsdii\ satisfies
Laplace's equation,
$$r^{-1}\ \partial_r\big(\,  r\partial_r T\, \big)+
{1\over 4\pi^2}\,
\partial_{\alpha}^2T=0\ .$$
This  allows one to calculate the second integral
in\ \uusldi. Thus we find,
$$M''_2=\int_r^{\infty} {dr\over 16\pi^2}
\, r\
\partial_{\alpha}^2T|_{
\alpha=0}\ \partial_{\alpha}^2T+
T|_{\alpha=0}+r\, \partial_r T|_{\alpha=0}\
\log\Big({4\pi N\over r}\Big)\ .$$
Finally we note that 
the most efficient way to study the\ $r\to 0$\ limit of
$M_2$\ \shdy\ is  based on the  representation\ \kksldoi.
It shows that for\ $r\to 0$\ with\ $r\cosh(\gamma)$\ fixed,
$${\cal Q}[\gamma]\big|_{r\to 0}\to
{\Gamma\big(1+
{r \cosh(\gamma)\over  
2\pi}\, \big)\over \sqrt{2r} \cosh(\gamma/2)}\
\Big({r \over 4\pi}\Big)^{ -{r \cosh(\gamma)
\over 2\pi}}\ \ \big(\, 1+O(b^2)\, \big)\ .$$
Hence, examination of  
the   product\ \shdy\ at this limit creates
no difficulties at all.

\listrefs

\end